\documentclass{nature}
\usepackage{epsfig,epsf,graphics,color,rotate,amsmath}
\usepackage{endnotes}
\let\footnote=\endnote

\title{Melting induced stratification above the Earth's inner core due to convective translation}

\author{Thierry Alboussi\`ere$^{1,2}$, Renaud Deguen$^1$ \&  Micka\"el Melzani$^1$ }

\begin{document}

\maketitle

\begin{affiliations}
\item Laboratoire de G\'eophysique Interne et Tectonophysique, CNRS, Observatoire de Grenoble, Universit\'e Joseph Fourier, Maison des G\'eosciences, BP 53, 38041 Grenoble Cedex 9, France
\item Universit\'e de Lyon, CNRS UMR5570, site UCB Lyon 1, 2 rue Rapha\"el Dubois, b\^atiment G\'eode, 69622 Villeurbanne, Universit\'e Lyon 1, ENS de Lyon, France
\end{affiliations}

\begin{abstract}
In addition to its global South-North anisotropy \cite{PoupinetPilletSouriau83}, there are two other major seismological observations relative to the Earth's inner core: asymmetry between the eastern and western hemispheres \cite{Tanaka97,Creager99,GarciaSouriau2000,Niu01,Yu05} and a layer of reduced seismic velocity at the base of the outer core \cite{SouriauPoupinet91,iasp91,Souriau95,ak135,PREM2,Yu05,Zou08}. This 250 km thick layer has been interpreted as a stably stratified region of reduced composition in light elements \cite{GubbinsMastersNimmo2008}. Here we show that this layer can be generated by simultaneous crystallization and melting at the surface of the inner core, and that a translational mode of thermal convection in the inner core can produce enough melting and crystallization on each hemisphere respectively for the dense layer to develop. The dynamical model we propose introduces a clear asymmetry between a melting and a crystallizing hemisphere which forms a basis for explaining the East-West asymmetry. The present translation rate is found to be typically 100~Ma for the inner core to be renewed entirely, which is one to two orders of magnitude faster than the growth rate of the inner core's radius. The resulting strong asymmetry of buoyancy flux caused by light elements is anticipated to have an impact on the dynamics of the outer core and on the geodynamo. 
\end{abstract}

The original observation \cite{SouriauPoupinet91} of seismic P-wave velocities slower than the adiabatic PREM model in the lower outer core has since been confirmed and incorporated in 1D global models AK135\cite{ak135} and PREM2\cite{PREM2}. That discrepancy from the adiabatic profile could result from a wrong interpretation due to the nearby complex inner core, as sensitivity kernels have a width of several hundreds of km at body waves frequency\cite{Calvet06}, or might also be attributed to floating crystals\cite{Loper1981,Zou08}. Gubbins {\it et al.} \cite{GubbinsMastersNimmo2008} show that this last explanation is not possible but that the observed seismic velocities can be explained by a stratification in light elements (and temperature). However, the stratification mechanism by crystallization and melting of crystals at different depths is not completely elucidated.

We propose that a dense layer can develop when melting and crystallization only occur at the inner core boundary (ICB). Where crystallization takes place, light elements are released providing light fluid and where melting takes place, dense fluid is produced. It is possible to quantify these effects in terms of flux of buoyancy. Let us denote $\Delta \rho $ that fraction of density jump across the ICB due to composition partition between solid and liquid phases. For a rate of crystallization $V$ (respectively melting), the buoyancy flux is $\Delta \rho  \ g_c \ V$ (respectively $-\Delta \rho  \ g_c \ V$), where $g_c$ is the magnitude of gravity \cite{PREM} on the ICB. The idea is that part of the heavy fluid would remain at the bottom, while the rest would be entrained by the light fluid. Conversely, part of the light fluid would mix with the dense fluid in the dense layer while the rest would cross the dense layer and contribute to convection within the main part of the outer core. This idea has been validated experimentally as follows.

The experiments consist in injecting simultaneously a constant flux of light and dense fluid at the bottom of a fluid cavity. The cavity is a box of perspex 20 cm high and with a 15 cm x 15 cm horizontal cross-section. It is initially filled with salted water ($c_0$~wt~\% NaCl). At the bottom of the cavity, there is a porous layer (sponge) below which the cross-section is divided into two disconnected parts: on one side light fluid is injected ($c_l < c_0$~wt~\% NaCl) and on the other side heavy fluid is injected ($c_h > c_0$~wt~\% NaCl). Both density differences $c_0 - c_l$ and $c_h - c_0$ and both flow rates are controlled and set constant during the experiment. The injections of fluids start simultaneously through pipes from reservoirs with the desired concentration. The excess of fluid is removed through an overflow at the top of the cavity. 

The geophysically relevant case is when the positive buoyancy flux exceeds the negative one since the inner core is growing on average. When the negative buoyancy flux induced by the heavy fluid is less than 80~\% in amplitude that of the light fluid, no dense layer is observed: the entrainment caused by the rise of light plumes is sufficient to mix the heavy fluid as it is released by the bottom boundary. 
However, when that heavy buoyancy flux is more than 80~\% that of the light buoyancy flux, a growing dense layer forms at the bottom of the cavity. It has been observed experimentally that the condition for the existence of the dense layer is really a condition on the buoyancy fluxes, as described above, and not a condition on the volume flow rates nor on the density differences between the fluids. This constitutes a justification for the relevance of such a convection experiment as a model of a melting/crystallization process for the inner core. 

On Fig. 1, an experimental run is shown. 
This experiment corresponds to a case where the heavy fluid buoyancy flux was 83~\% that of the light fluid. The initial concentration and concentrations of the dense and light injected fluids were $c_0 = 4$~wt~\%, $ c_h = 6$~wt~\% and $c_l = 1.65$~ wt~\% NaCl respectively. The volume flow rate of the dense fluid was $Q_h = 3.9 \ 10^{-7}$m$^3$s$^{-1}$ and that of the light fluid $Q_l = 4.0 \ 10^{-7}$m$^3$s$^{-1}$. 
That experiment has been run twice under the same conditions: in the first instance, the injected dense fluid was colored with potassium permanganate and photographs of the setup have been taken at different times after the beginning of the injections. A dense colored layer forms at the bottom and its thickness grows linearly with time. It is also possible to see convection plumes going up on the right-hand side, carrying along some of the heavy colored fluid in the upper part of the cavity. In the second instance, the synthetic schlieren method has been used \cite{DalzielHughesSutherland2000,GostiauxDauxois2007}, providing a quantitative two-dimensional field of refraction index, hence visualizing concentration gradients: their horizontal components are shown on the middle row of Fig. 1 showing convection plumes of light fluid on the right-hand side of the cavity,  while their vertical components are shown on the bottom row visualizing the dense layer and its growth. The concentration field is computed from its gradient, and averaged along the horizontal direction: the resulting stratification profile is shown on Fig. 2. There is clearly a region of stratified fluid, above which density is nearly uniform. The thickness of this layer is growing linearly with time, its volume being 50~\% to 90~\% that of the total volume generated by the light and heavy fluxes.   

Melting part of the inner core at a significant rate is difficult while it is crystallizing on average as a result of secular cooling. The most plausible way is that a topography is formed dynamically on the ICB so that the temperature of adjacent fluid of the outer core exceeds the melting temperature. That excess temperature is then responsible for heat transfer from the outer core to the ICB, providing latent heat for fusion: hence topography can be related to the rate of melting. 

The dynamical model we put forward to account for significant melting on the ICB results from the combination of three physical elements: thermal state of a superadiabatic inner core, gravitational equilibrium and finite heat exchange of latent heat with the outer core. In superadiabatic conditions, a uniform velocity in the inner core $V$, say from West to East along the $x$-axis (see Fig.~3), generates a global superadiabatic temperature gradient in the same direction proportional to the residence time in the inner core, hence inversely proportional to $V$, and proportional to a positive source term $S \sim 10^{-15}$~K~s$^{-1}$ defined from secular cooling and thermal conduction along the adiabat (see Methods Section and reference \cite{StaceyDavisBook2008}):
\begin{equation}
\frac{\partial \Theta }{\partial x } = \frac{S}{V},\label{gradient_temperature}
\end{equation}
where $\Theta$ is the temperature relative to the adiabat $T_{ad}$ in the inner core anchored to the ICB \cite{PREM}. 
It follows from the volume expansion coefficient \cite{Vocadlo07} $\alpha = 1.1\ 10^{-5}\,$K$^{-1}$ and inner core density (on the ICB \cite{PREM}) $\rho _s = 1.28\ 10^4\ $kg$\, $m$^{-3}$, that there exists a density gradient $- \alpha \rho _s {\partial \Theta }/{\partial x }$. The resulting gravity field and density distribution generate unbalanced forces on the inner core, so that it is displaced a distance $\delta$ in the $x$-direction. 
In the Methods section, the gravitational field and potential associated with this mass distribution are derived, from which it is possible to calculate the net gravitational force ${\bf F}_G$ exerted on the inner core and the net pressure force ${\bf F}_P$ exerted by the outer core on the inner core: 
\begin{equation}
{\bf F}_G + {\bf F}_P = \frac{16 \ \pi ^2}{9} {\cal{G}}\ \rho _\ell c^3 \left[ \alpha \frac{\partial \Theta }{\partial x } \rho _s \frac{c^2}{5}  - ( \rho _s - \rho _\ell ) \delta \right]  \ {\bf e}_x , \label{gravity_forces_linear}
\end{equation}
where ${\cal{G}}$ is the universal gravitational constant, $\rho _\ell$ is the outer core density on the ICB \cite{PREM} and ${\bf e}_x$ is the unit vector in the direction of the temperature gradient. 
The equilibrium condition that both forces balance provides the shift $\delta$ as a function of the thermal gradient $\partial \Theta / \partial x$: 
\begin{equation}
\delta = \frac{\alpha \ \frac{\partial \Theta}{\partial x} \rho _s c^2}{5 \ (\rho _s - \rho _\ell )}. \label{equilibrium}
\end{equation}
Then, the displacement $\delta$ is associated with a non uniform pressure distribution on the ICB (see Methods Section) hence to a small temperature departure $\delta T$ from the adiabat (see Fig.~4): 
\begin{equation}
\delta T = \rho_\ell g_c \ \delta \ \cos \theta  \ (m_P - m_{ad}) , \label{temperature_variation}
\end{equation}
where $m_P = 8.5\ 10^{-9}$~K~Pa$^{-1}$ is the Clapeyron slope \cite{Alfe02b}, $m_{ad} = (\alpha T_{ad})/( \rho c_p ) = 6\ 10^{-9}$~K~Pa$^{-1}$ is the adiabatic gradient and $g_c = \mathcal{G} \frac{4 \pi}{3}\rho_s c$ is gravity on the ICB. 
That departure is accomodated by a thermal boundary layer in the outer core, with a corresponding heat transfer of typical magnitude $u'\, c_p \, \delta T$, where $u' = 10^{-4}$~m~s$^{-1}$ is a typical velocity scale in the outer core and $c_p = 850$~J~kg$^{-1}$~K$^{-1}$ is the specific heat capacity \cite{Poirier1994a}. That heat transfer must be balanced by the release or absorption of latent heat:
\begin{equation}
L \ V \ \cos \theta = u'\ c_p\ \delta T,   \label{heat_transfer}
\end{equation}
where $L\, = \, 900$~kJ~kg$^{-1}$ is the latent heat coefficient \cite{Poirier93,Anderson1997}. 
Finally, combining equations (\ref{gradient_temperature}), (\ref{equilibrium}), (\ref{temperature_variation}) and (\ref{heat_transfer}), one can express the translational velocity:
\begin{equation}
V^2 = \frac{4 \pi \mathcal{G}}{15}\frac{ u' \ c_p \ \rho _s^2 \rho_\ell \ \alpha  \ (m_P - m_{ad} ) \ S }{L \ (\rho _s - \rho _\ell )}\ c^3 . \label{solutionV}
\end{equation}

Depending on the heat flux at the CMB, the history of the inner core shows a first phase dominated by growth $\dot{c} \sim c^{-1}$, followed by the development of the translational instability (see Supplementary information) when its radius was around 400~km, leading to the dominant present translation $V\sim c^{3/2}$ of order $5 \ 10^{-10}$~m~s$^{-1}$ while the growth rate is of order $10^{-11}$~m~s$^{-1}$ (Fig.~5).  

The latter scaling law implies that the translational convection is faster along a long axis of the inner core oblate spheroid (see Supplementary information), {\it i.e.} perpendicular to the rotation axis. It follows that the temperature gradient is preferrentially aligned with such a long axis, which iagain reinforces convection in that direction. Moreover, the Earth's aspherical mass distribution which has essentially a degree 2, order 2 geometry \cite{MJSG82} is responsible for elongating slightly the inner core along an East-West axis and induces a degree 1 translational convection in the inner core through a bifurcation produced by instability (see supplementary information). We propose that the translational flow has a West to East orientation, hence being responsible for the observed hemispherical asymmetry of the inner core: grain growth during the transit from the western hemisphere to the eastern hemisphere may explain the difference in seismic properties \cite{CalvetMargerin08}. The temperature difference of a few K between both hemispheres is another source of asymmetry.

According to our experiments, a melting rate above 80~\% that of the crystallization rate is necessary for a dense layer to form, which geometrically implies that the translation velocity $V$ is more than 20 times that of the inner core growth rate. From Fig.~5, this happens only when the CMB heat flux exceeds 10~TW, and only since the inner core radius was 1100~km, some 200~Ma ago. Extrapolating from our experiments, evaluating 50~\% of the volume of melt produced since then would corresponds to a layer of thickness 250~km. The experimental excess concentration is found to be 10~\% of the concentration difference between light and heavy injected fluids. In the Earth's core where concentration in light element is of order 10~\%, a concentration difference around 1~\% across the dense layer is expected. This is indeed coherent with the observed seismic velocities \cite{GubbinsMastersNimmo2008}.

Our convection mechanism ignores deformation in the inner core and compositional buoyancy. With a finite effective viscosity, temperature variations along gravity isopotentials induce an internal flow with deformation that affects the translational mode. We have estimated that an internal flow is weak compared to translation for an effective viscosity above $10^{18} \ $Pa~s. Enrichment in light elements of the outer core (a few percents) has been invoked \cite{buffett09,DeguenCardin09} to imagine a stabilizing mechanism for convection in the inner core. This is however very speculative, as the fraction of light elements incorporated in the inner core may have decreased more rapidly than the increase in outer core composition since gravity on the ICB is getting larger, reinforcing compaction. 

Invoking an excessively asymmetric buoyancy flux on the ICB deserves further study relative to the dynamics of the outer core and the geodynamo. The stratified layer is expected to be dynamically isolated and to act as a filter between the inner core and the rest of the outer core, but there might subsist some hemispherical asymmetry in the outer core dynamics.

\begin{figure}
\includegraphics[width=15cm, keepaspectratio]{figure1.pdf}
\caption{{\bf Visualization of the growth of a dense layer in an experimental run, using (a) dye injection and (b) horizontal and (c) vertical density gradients.} The experimental cavity is initially filled with a 4 wt\% NaCl water solution. From $t=0$, a constant flux of 1.65 wt\% NaCl solution is injected at the bottom on the right-hand side while a 6 wt\% NaCl solution is injected on the left-hand side. The dense fluid is colored with permanganate potassium on the top row, visualizing a growing dense layer at the bottom, at four different times after the beginning of the injection. The synthetic schlieren method is used in a second identical experiment: the horizontal gradient of refraction index on the middle row highlights the convective plumes and the vertical gradient on the bottom row reveals the dense layer. }
\end{figure}

\begin{figure}
\includegraphics[width=15cm, keepaspectratio]{figure2.pdf}
\caption{{\bf Evolution of the concentration profile during the growth of a dense layer.} The concentration field is extracted from the gradient of refraction index. It is averaged along the horizontal direction and shows the evolution of the dense layer. }
\label{sketch_model}
\end{figure}

\begin{figure}
\includegraphics[width=15cm, keepaspectratio]{figure3.pdf}
\caption{{\bf A schematic representation of the translational convective mode.} The centre of the inner core O, is shifted by a distance $\delta$ away from the centre of the Earth $C$, which would be its equilibrium position if its density were uniform. That shift causes thermodynamic disequilibrium at the ICB, generating melting on one side and crystallization on the other one. Hence a uniform flow exists in the inner core: in the case of a superadiabatic regime, a gradient of temperature develops as represented in grey scale. Its associated changes in density and gravitational potential lead to a new mechanical equilibrium for the inner core, corresponding indeed to a shift in position in the same direction as initially assumed.  }
\end{figure}

\begin{figure}
\includegraphics[width=15cm, keepaspectratio]{figure4.pdf}
\caption{{\bf Thermal departure from the adiabat due to the displacement of the inner core and heat transfer at the ICB.} A thermal boundary layer forms in the outer core to adjust to the different radii of the ICB on the melting and crystallization sides.}
\end{figure}

\begin{figure}
\includegraphics[width=15cm, keepaspectratio]{figure5.pdf}
\caption{
{\bf Growth rate of the radius of the inner core and uniform convective velocity as functions of the inner core radius.} They are plotted for different values of the heat flux at the CMB. Light solid lines : mean solidifcation rate of the inner core. Light dash-dotted lines : translation velocity, calculated with the assumption of a constant $S$. Heavy solid lines : translation velocity, with $S(t)$ calculated (see supplementary information) from the core thermal evolution model of\cite{Labrosse2003}.
}
\end{figure}

\vfill

\noindent
{\bf METHODS SUMMARY}\\
The mode of convection associated to the translation of the inner core is not standard. Therefore, it is presented in the Methods section. Thermal buoyancy is the driving force, however, unlike classical convection, the damping is not due to viscous and/or thermal diffusion. Damping is set by the capacity of the outer core to extract or supply latent heat on the ICB. 

\bibliography{bib.bib}

\noindent
{\bf Supplementary information} is linked to the online version of the paper at www.nature.com/nature

\noindent
{\bf Acknowledgements} This work has benefited from fruitful discussions during the CNRS-INSU SEDIT meetings. We thank M. Bergman for discussions regarding inner core crystallization. The LGIT and the ANR (Agence Nationale de la Recherche) have provided financial support for the experiments. 

\noindent
{\bf Author contributions} M.M., R.D. and T.A. ran and analyzed the experiments; T.A. designed the experimental study and built the dynamical model; R.D. and T.A. worked out the thermal conditions on the ICB and assessed the geophysical relevance of the dynamical model; R.D. computed the different scenarios of thermal history; R.D., T.A. and M.M. applied the experimental results to the geophysical context; T.A. and R.D. wrote the paper. 

\noindent
{\bf Author information} Reprints and permissions information is available at www.nature.com/reprints. Correspondence and requests for materials should be addressed to T.A. (thierry.alboussiere@ens-lyon.fr).

\end{document}